\documentclass[preprint,12pt]{elsarticle}

\usepackage[most]{tcolorbox}
\usepackage{rotating}
\usepackage{soul}

\usepackage{hyperref}
\usepackage{listings}
\usepackage{array}
\usepackage{longtable}
\newcolumntype{N}{>{\centering\arraybackslash}m{.5in}}
\newcolumntype{G}{>{\centering\arraybackslash}m{2in}}
\usepackage{subfig}
\usepackage{mathtools}
\usepackage{algorithm}
\usepackage{balance}
\makeatletter
\usepackage{multirow}
\def\BState{\State\hskip-\ALG@thistlm}
\makeatother
\usepackage{booktabs}

\usepackage{algorithmic}
\usepackage{graphicx}
\usepackage{textcomp}
\usepackage{xcolor}

\hypersetup{
	colorlinks   = true, %Colours links instead of ugly boxes
	urlcolor     = blue, %Colour for external hyperlinks
	linkcolor    = blue, %Colour of internal links
	citecolor   = blue %Colour of citations
}

\usepackage{amssymb}
\usepackage{color}
\journal{Nuclear Physics B}

\begin{document}

\begin{frontmatter}



\title{Dev2vec: Representing Domain Expertise of Developers in an Embedding Space}


\author[inst1]{Arghavan Moradi Dakhel}

\affiliation[inst1]{organization={Department of Computer and Software Engineering},%Department and Organization
            addressline={Polytechnique Montreal}, 
            city={Montreal},
            postcode={H3T 1J4},
            state={Quebec},
            country={Canada}}

\author[inst1]{Michel C. Desmarais}
\author[inst1]{Foutse Khomh}

\begin{abstract}
Accurate assessment of the domain expertise of developers is important for assigning the proper candidate to contribute to a project, or to attend a job role. Since the potential candidate can come from a large pool, the automated assessment of this domain expertise is a desirable goal. While previous methods have had some success within a single software project, the assessment of a developer's domain expertise from contributions across multiple projects is more challenging.

In this paper, we employ doc2vec to represent the domain expertise of developers as embedding vectors. These vectors are derived from different sources that contain evidence of developers' expertise, such as the description of repositories that they contributed, their issue resolving history, and API calls in their commits. We name it dev2vec and demonstrate its effectiveness in representing the technical specialization of developers.

Our results indicate that encoding the expertise of developers in an embedding vector outperforms state-of-the-art methods and improves the F1-score up to 21\%. Moreover, our findings suggest that ``issue resolving history'' of developers is the most informative source of information to represent the domain expertise of developers in embedding spaces.

\end{abstract}




\end{frontmatter}


\section{Introduction}
\label{sec:intro}
 %appendix~\ref{sec:sample:appendix}.
Large software projects require technical experts in different domains~\cite{curtis1988field}. Accurate assessment of developers' expertise impacts the success of software projects~\cite{demarco2013peopleware}. This assessment includes details of their soft skills, such as communication skills~\cite{liang2022towards,zhou2018makes} and technical skills, such as programming languages or libraries~\cite{matter2009assigning, montandon2019identifying}. Automatic identification of the technical expertise of developers in different domains is a significant challenge in software engineering because the appropriate candidate is generally selected from a large pool.

Platforms such as GitHub gather useful information from a large pool of programmers that an be mined to extract the domain expertise of developers based on their activities~\cite{marlow2013activity}. These platforms include the activity of developers over different open source repositories. However, GitHub is a version controlling system and is not an appropriate tool for specifically searching and filtering experts. A survey~\cite{singer2013mutual} shows recruiters use platforms such as GitHub to search for an expert and 66\% of non-technical recruiters struggled with technical information contained in these platforms. Therefore, methods to process the information in such platforms and assess the domain expertise of developers are required.

Some studies have relied on general heuristics such as Line 10 rule and count the number of contributions of developers on different source files to find experts~\cite{mcdonald2000expertise,mockus2002expertise}. Other studies present more specific methods to identify an expert to perform a task, fix a bug, or use a library. They count the frequency of API calls in their commits~\cite{montandon2019identifying,oliveira2019well,schuler2008mining} or use methods such as bag-of-words on textual information available to infer the expertise of developers~\cite{greene2016cvexplorer,matter2009assigning}.

While these methods show good performance in identifying an expert within a software project, they suffer from data sparsity and high dimensionality when used to represent the domain expertise of developers across different types of activities and different software projects~\cite{montandon2021mining}. Also, these methods struggle to extract the semantics of words.
%and their order in the text. There are methods such as bag-of-n-grams that considers the order between words~\cite{arora2018compressed} but bag-of-n-grams incur higher dimension and cause more sparsity. 

One of the solutions to extend the semantic capabilities of these methods is to rely on embeddings. Words, or aggregation of words, are encoded as fixed-length vectors, resulting in low dimensionality representation. \verb|word2vec|~\cite{church2017word2vec} and \verb|doc2vec|~\cite{le2014distributed} are well known examples. These methods are known to improve the performance of different learning tasks in natural language processing such as classification~\cite{ge2017improving}. A recent study~\cite{dey2021representation} recommends new contributors to a software project by applying doc2vec on the list of APIs in source files that developers changed in different software projects in the past.

In this paper, we demonstrate a suitable presentation for the domain expertise of developers. It captures semantic similarity within different domains of expertise. We address the problem of identifying the domain expertise of developers across their activities in various software projects on GitHub. 
We employ doc2vec to encode the domain expertise of developers in embedding vectors.

There are different types of information about developers' activities on GitHub. These sources of information can inform the representation of developers' expertise. We collect three different sources of information from GitHub that are prominent sources to assess the domain expertise of developers~\cite{matter2009assigning,greene2016cvexplorer,wan2018scsminer}. We derive embedding vectors of developers' expertise from the meta-data of repositories that developers contributed to, the issue resolving history of developers and the list of APIs in changes applied by developers on different source files.  We name these methods after their respective sources: \textit{dev2vec:Repos}, \textit{dev2vec:Issues} and \textit{dev2vec:APIs}, respectively. In addition, we merge the output of these three methods by concatenating the embedding vectors from these three different spaces of information and call it \textit{dev2vec:RIAs} (RIA stands for \textbf{R}epository, \textbf{I}ssue and \textbf{A}PI). 

We evaluate the performance of each dev2vec model in reflecting the domain expertise of developers over a favored task in software engineering: identifying the technical job roles of developers. We use a labeled dataset of GitHub developers. They are categorized into five job roles: Backend, Frontend, Mobile, DevOps, and Data Scientist~\cite{montandon2021mining}. Specifically, we answer three different research questions in this work:\\
\textbf{RQ1:} How effective are embedding vectors to represent the domain expertise of developers across various software projects compared to the state-of-the-art methods? \\
\textbf{RQ2:} How sensitive is the performance of dev2vec to the source of information (repositories, issues and API)?\\
\textbf{RQ3:} How effective is the concatenation of expertise embedding vectors from the different information sources?\\

% to represent the expertise of developers

Our experimental results show that encoding the expertise of developers in an embedding vector outperforms a state-of-the-art study~\cite{montandon2021mining} that used bag-of-words with dimension reduction techniques on GitHub information to represent the expertise of developers. 
%\Foutse{when people read the paper 10 years from now, they wont know what recent means....i think it is better to just say that you outperform the state of the art technique X (adding the reference!)} 
Also, the performance of dev2vec is sensitive to the sources of information: issue resolving history of developers results in better performance.

This paper makes the following contributions:
\begin{itemize}
%\item Predict embedding vectors that represent the expertise of developers by concatenating APIs embedded vectors derived from a pre-trained model on API-related skill space of developers (\textit{dev2vec:APIs}).
\item Propose a method to represent the expertise of developers in embedding vectors by applying doc2vec on different sources of information: \textit{dev2vec:Repos}, \textit{dev2vec:Issues}, and \textit{dev2vec:APIs}.

\item Propose to aggregate three different types of information about developers' activities across different projects on GitHub and represent the domain expertise of developers by concatenating the embedding vectors of developers' expertise from different spaces: \textit{dev2vec:RIAs}.

\item Evaluate the effectiveness of dev2vec to represent the domain expertise of developers by a prevalent problem in the software engineering domain: assessing the technical specialization of developers.

\end{itemize}

%\Foutse{if you have space, you can add a paragraph describing the outline of the paper!}
%\Foutse{i think we have space for this, we are only at 9 pages!}
\textbf{The rest of this paper is organized as follows.} Section~\ref{sec:DtCollection} explains the ground truth and describes different datasets used in experiments. Section~\ref{sec:method} describes the methodology of our study. Section~\ref{eval} presents the baseline and experimental setup.  Section~\ref{sec:TV} highlights different factors that constitute threats to the validity of our experiment. In Section~\ref{disc}, we discuss our findings and their application in other studies. Section~\ref{sec:RW} introduces the related literature. Finally, Section~\ref{CL} concludes the paper and discusses avenues for future work.
%%%%%%%%%%%%%%%%%%%%%%%%%%%%%%%%%%%%%%%%%%%%%%%

%%%%%%%%%%%%%%%%%%%%%%%%%%%%%%%%%%
\section{Data collection} \label{sec:DtCollection}
\begin{figure*}
  \includegraphics[scale=0.5]{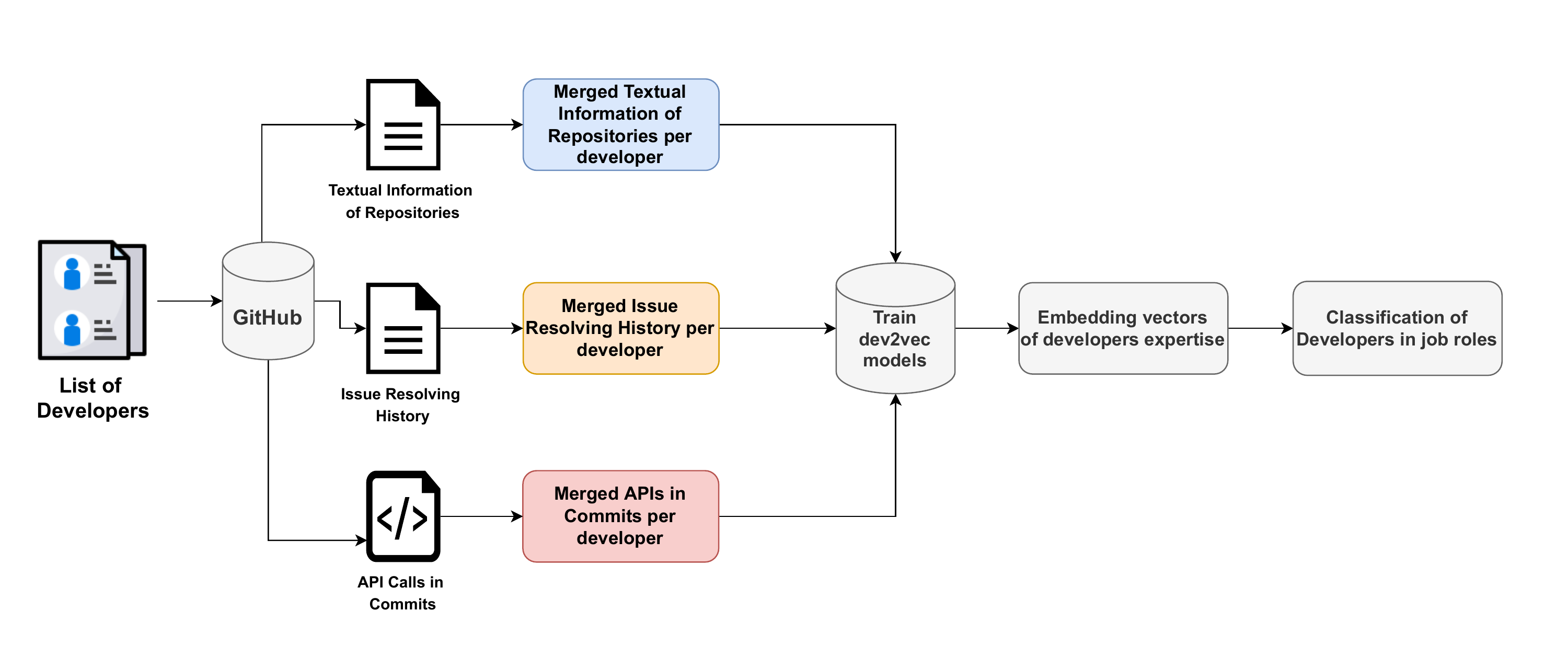}
  \caption{\textbf{Overview of the pipeline.} This is an overview of the proposed methods. Three separate sources of information are collected to represent developer expertise as embedding vectors.}
  \label{fig:overall}
\end{figure*}

To determine the domain expertise of developers, we rely on three different types of information %\Foutse{you are not listing activities but information types} 
available from GitHub. The first one is the information of repositories %\Foutse{this is not an activity! you mentioned 3 activities...an information is not an activity...please rephrased} 
that they contributed to %\Foutse{i would avoid having 'to on' it is akward!} 
GitHub. The second is their issue resolving history %\Foutse{again this is not an activity!} 
on different repositories. The last is the list of APIs in changes applied by them across different repositories. Figure~\ref{fig:overall} shows an overview of the data collection process. In the rest of this section, we first introduce how we obtain the ground truth labeling. Then, we explain how to collect the data for these three categories.

\subsection{Ground Truth}

Our training and validation method relies on the availability of labeled data that represents the ground truth developer expertise.  We used a labeled dataset of developers published in \cite{montandon2021mining}. It includes 1662 developers, categorized into five job roles: \textit{Backend, Frontend, Mobile, DevOps, and Data Scientist}. They did not use GitHub's data to generate expertise labels. Instead, they used the StackOverflow data of its users with GitHub profiles. They extracted users' job roles by applying different regular expressions to the profile information of their StackOverflow. Also, they collected the GitHub username of these developers to link them to their GitHub pages.

\begin{figure}
\centering
  \includegraphics[width=0.6\textwidth]{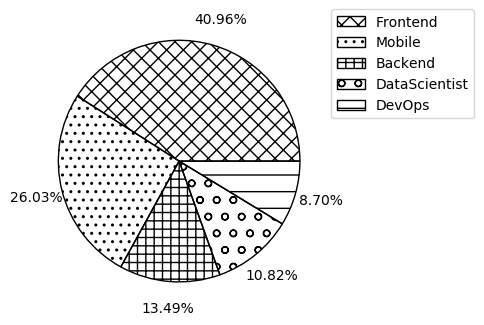}
  \caption{\textbf{The distribution of developers in different job roles. %\Foutse{can you use shapes instead of colors?}
  } The distribution of 1272 developers in five job roles: Backend, Frontend, Mobile, DevOps, and Data Scientist. This distribution is imbalanced.}
  \label{fig:jobroles}
\end{figure}

20\% of these developers have more than one label. We limit the scope of this study to developers with single domain expertise and kept 1340 developers with a single job role. Out of these 1340 developers, we successfully find the GitHub profile of 1272 developers with their usernames. The remaining 68 developers may have changed their usernames or closed their accounts on GitHub. Figure~\ref{fig:jobroles} shows the distribution of developers in these five job categories.

%\begin{figure}
  %\includegraphics[width=0.45\textwidth]{jobroles.jpg}
  %\caption{The distribution of developers in 5 job roles}
  %\label{fig:jobroles}
%\end{figure}

\subsection{Textual Information of Repositories} \label{subsec:repos}
Based on previous studies, information about the repositories that developers contributed in the past is a good reference to determine their skills and domain of expertise~\cite{greene2016cvexplorer,wan2018scsminer}. 

In GitHub, each repository has different features in natural language. These features can describe its purpose of the domain. They are included \textit{Name, Tags, Topics and ReadMe}. ``Name'' is the repository's title, and it is mandatory. Repository names can be chosen in a way to represent the goal of the project to some extent. However, these names are short, and are not very descriptive (e.g.~``eth-tester-rpc~\footnote{\url{https://github.com/voith/eth-tester-rpc}}''). ``Tags'' can be assigned to a repository to specify its domain, programming language or technologies are used in the project (i.e. \#ethereum, \#python, \#crypto). Tags are not mandatory and can be left empty. ``Topic'' or ``About'' is a short description, up to two or three sentences, about the goal of the repository. The ``ReadMe'' file describes more details such as the application of the project, the functionality of different source files, programming languages used inside the repository, etc. Topic and ReadMe are optional, too. 

We combine the content of all these four features for each repository and call it the textual information of repositories. Then, we collect the textual information of all repositories and aggregate it on a per developer basis. The output is a document per developer that contains the textual information from repositories they contributed.  If a repository is forked, we find the parent repository and, if the developer has any participation (commit) on the parent repository, we collect its description too. Overall, we collect the textual information of 58K repositories for all 1272 developers.

\subsection{Issue Resolving History}\label{sec:issue} 

Issue resolving history of developers across different projects on GitHub is another source that contain useful information about their domain expertise~\cite{tian2016learning,matter2009assigning}. Developers can have different types of contributions on an issue in GitHub repositories. An issue can be assigned to a developer or a list of developers. It can be created by a developer on a repository. Alternatively, a developer can participate in the discussion of an issue. We consider all these cases as a contribution of developers to solving an issue on GitHub and call it the Issue Resolving History of developers. The mixture of these activities can give us valuable information about developers' expertise.

We collect the header and the body of 60K issues assigned to 1272 developers. We combine the header and the body of each issue as its content. Then for each developer, we merge the content of all her/his issue resolving history in the past. Akin to the textual information of repositories, the output of this step is a document per developer, and each document contains the issue resolving history of a developer.

\subsection{API Calls per Commits} 
The main contributions of developers on different repositories are in the format of commits. Another useful piece of information that is considered as evidence of developers' expertise is the list of API calls in commits across different repositories on GitHub.~\cite{schuler2008mining,venkataramani2013discovery}. Developers change one or more source file(s) by submitting a commit. Each source file contains a list of APIs that developer may or may not change. Based on previous studies, we assume that developers have a basic knowledge about the APIs used in the source files they modified~\cite{dey2021representation}. 

For this step, per each commit, we extract all language-specific source files linked to a commit after submitting it. Overall, we extract source files of 21M commits (the changed version of the source file after a commit). As a result, for each developer, we merge all APIs in the source files of their commits.  In section~\ref{sec:dev2vec:APIs}, we explain more details about the process of API collection of developers' commits. 

%%%%%%%%%%%%%%%%%%%%%%%%%%%%%%%%%%%%%%%%%
\section{Deriving Dev2vec from doc2vec} \label{sec:method}
In this section, we describe how to represent the expertise of developers in embedding vectors with three proposed methods, dev2vec:Repos, dev2vec:Issues and dev2vec:APIs, and describe how to train different the models with the data sources from Section~\ref{sec:DtCollection}, and how .  Also, we explain \textit{dev2vec:RIAs}, the method for combining these three embedding vectors of expertise. 

%Finally, we illustrate how to evaluate the efficiency of these vectors in representing the domain expertise of developers.

\begin{figure*}
  \includegraphics[scale=0.4]{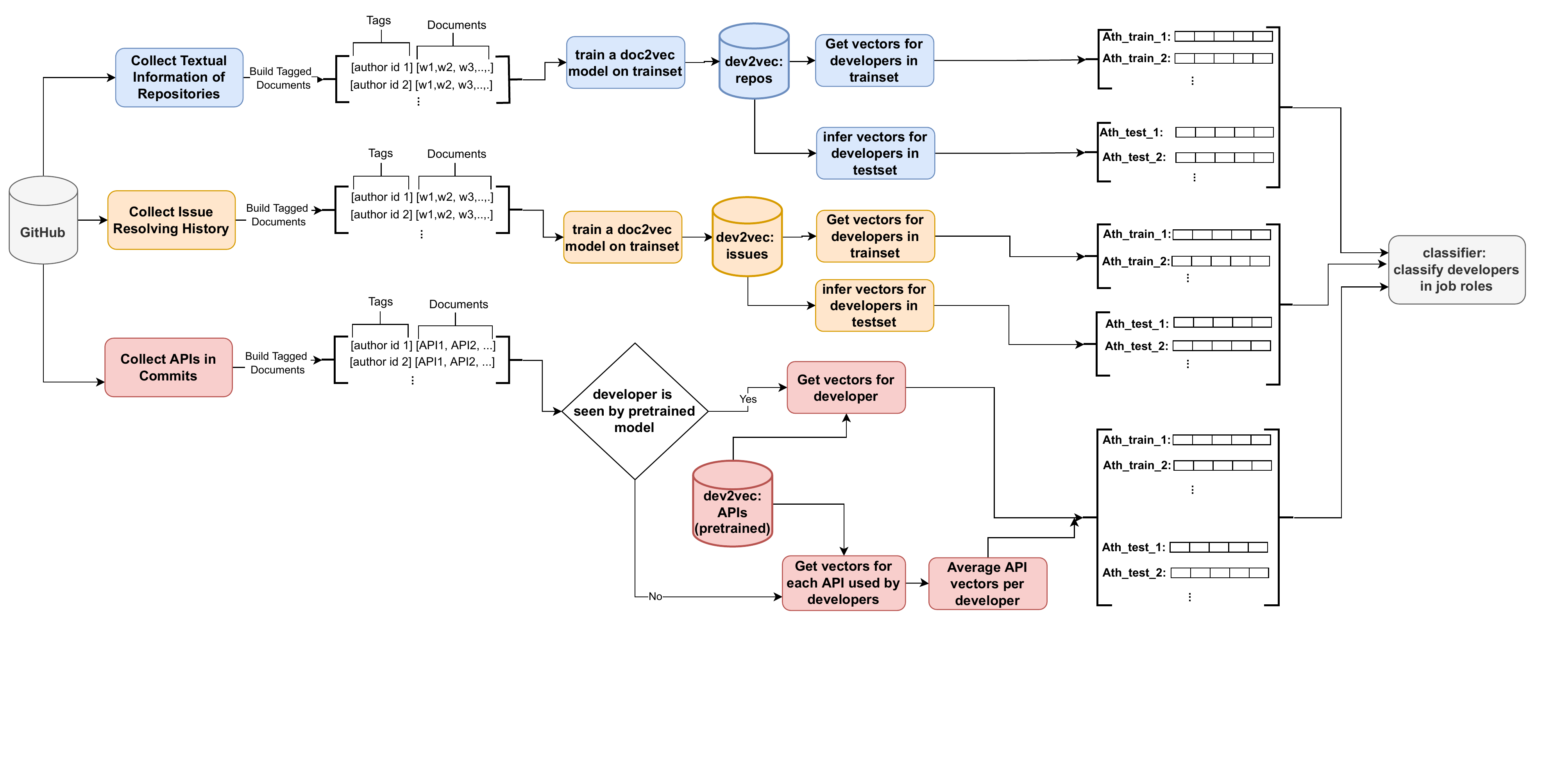}
  \caption{\textbf{Detailed view of the pipeline.} A more detailed view of the three proposed methods, dev2vec:Repos, dev2vec:Issues and dev2vec:APIs. The embedding vectors generated by each of these models represent the expertise of developers in different embedding spaces. Each category of embedding vectors is fed into a classifier separately to evaluate the their effectiveness in representing developer's expertise}
  \label{fig:approach}
\end{figure*}

\subsection{The embedding of domain expertise with doc2vec}
Assessing the expertise of developers across different software projects and different programming languages, regardless of the data source, increases the number of distinct tokens. For example, after cleaning the description of repositories, we end up with 18M distinct keywords. The dimension increases to 100M-dimension for API calls~\cite{dey2021representation}. Previous methods that show good performance to find an expert within a single software project, such as bag-of-words method, are not practical with these high dimensions. 

As we already discussed, one of the solutions to address these kind of problems is to encode this information into fixed-length vectors with much lower dimensions by word2vec or doc2vec algorithms. word2vec~\cite{church2017word2vec} learns a numerical vector representation for each word in a corpus of documents with two different algorithms CBOW and skip-gram. The vector size is much less than the vocabulary dimension, mostly between 100 to 300~\cite{church2017word2vec}. 

An extension of word2vec is doc2vec. It represents each document in an embedding vector and is trained based on predicting the words of a document within a corpus. The text that we feed into doc2vec can have variable-length ranging from a single sentence to a large document~\cite{le2014distributed}.  In doc2vec, each document has a paragraph id or tag. This tag can be a single identifier or a list of identifiers per document. Doc2vec learns an embedding vector for each identifier of documents and each word of documents. This embedding vector is a representation of the document. 

%In this case, doc2vec learns a vector for this tag based on the content of all documents that shared same tag.

The Doc2vec algorithm has two stages. The first stage is the training. There are two approaches to training document vectors and word vectors. The first one is: Distributed Memory model (DM). In this approach document embedding vector acts as an additional input word, and the average (or concatenation) of its vector with the embedding vectors of the rest of input words predicts output word. The embedding vector of documents are considered as a memory of the document content. The second approach is: Distributed Bag Of Words (DBOW). In DBOW, the document embedding vector is learned to predict randomly sampled words in output~\cite{le2014distributed}.

The second stage is the inference phase (or predicting phase).   In inferring stage, the vectors of new documents are inferred by concatenating their word vectors and repeating several iterations until convergence. Words that were unseen in training are ignored.

\subsection{dev2vec:Repos}\label{sec:repos}
The dev2vec:Repos model uses the data collected in %\Foutse{when you refer to a specific section it should be a capital letter 'Section'...please fix other places!}
Section~\ref{subsec:repos} to train a doc2vec model. We consider the textual information of repositories per developer as a document. After tokenization and cleaning, we build tagged-document data by tagging each document by the username of its developer or Author id. The first pipeline in Figure~\ref{fig:approach} shows the data entries that we use to build \textit{dev2vec:Repos}. It includes \textit{[developer id]} as tag or paragraph id and \textit{[w1,w2,...,wn]} as document content. \textit{[w1, w2, ..., wn]} are words collected from the description of repositories that a developer contributed in the past. 

We divide developers into traintet and testset. We train the doc2vec model on tagged documents of developers in trainset and infer vectors for developers in testset based on the content of their documents. It is worth mentioning that if there is a new word in textual information of repositories for a developer in testset (the word is not seen during the trainset), the inference stage ignores that new word. In Section~\ref{sec:setup}, we provide more details on the experimental setup of this model.

\subsection{dev2vec:Issues}
The dev2vec:Issues model uses the data described in Section~\ref{sec:issue}. Akin to the approach in Section~\ref{sec:repos},  we consider the issue resolving history of each developer as a document. We tokenize and clean the content of each document. Then, we tag each document by its related developer's username or Author id to build the tagged documents. The second pipeline in Figure~\ref{fig:approach} shows the format of data to train \textit{dev2vec:Issues}. It includes \textit{[developer id]} as the tag and \textit{[w1,w2,...,wn]} as document content. \textit{[w1, w2, ..., wn]} are words collected from the title and body of issues that a developer solved, created, or contributed in the past. 

In the training phase, the doc2vec model learns a vector for each developer in the trainset. Then, we infer or predict vectors for developers in the testset based on this new model. Akin to the inference stage in Section~\ref{sec:repos}, if a new word appears in issue resolving history of a developer in testset, it will be ignored. Section~\ref{sec:setup} provides  more details on the experimental setup of this model.

\subsection{dev2vec:APIs} \label{sec:dev2vec:APIs}
The dev2vec:APIs model uses a pre-trained doc2vec model~\cite{dey2021representation}, which is trained on API calls in source files collected from GitHub. We chose this model because it is trained on a vast dataset of 36K projects, 690K authors, and 1.2B source files after submitting commits. Training such a huge model is time-consuming and needs high-capacity infrastructure. In addition, we cannot infer the embedding vector for new developers using this model directly  (not seen in their training phase) due to the granularity level of data entries in their model. Therefore we introduce a new method with {dev2vec:APIs} to use this model for predicting the embedding vector of developers' expertise based on their API usage. It allows us to indirectly compare this model with {dev2vec:Repos} and {dev2vec:Issues}. In the following section, we provide more details on different parts of {dev2vec:APIs}.

\subsubsection{Parsing source files}  Since we use the pre-trained model in~\cite{dey2021representation}, we follow the same structure to collect the API calls in the source files. They collected the list of libraries (imports) in the version of source files after submitting a commit for 17 programming languages: C, C\#, Java, FORTRAN, Go, JavaScript, Python, R, Rust, Scala, Perl, Ruby, Dart, Kotlin, TypeScript, Julia and Jupyter Notebook (iPython). We parse each source file based on the syntax of its programming language and collect the list of its libraries for each developer. We match the extension of each source file after submitting a commit with its programming language. Then, based on the grammar of its programming language, we apply the related regular expression to collect its list of libraries. For example, for Python, we use this regular expression to find libraries:  ``$(?:from\vert import)\backslash s+\backslash w*. +;*$''.

\subsubsection{Obtaining embedding vectors  of developers expertise} 

In the pre-trained model~\cite{dey2021representation} the granularity level of the input data is fed into the doc2vec model, is per commit (changed version of source file after submitting a commit). The model used a list of identifiers: \textit{[programming language, repository, timestamp, author id]} as the document tag and the list of APIs in the changed version of the source file as the content of the document or words. The Author id belongs to the developer who modified this source file by submitting the commit. 

In the inference stage of doc2vec, we must infer document vectors for the same granularity level of documents as in the trainset. However, we can get or fetch the vector of documents and words which have been seen and trained during training. Of all 1272 developers in our dataset, only 40 of them are in their dataset. We directly extract the embedding vectors of these 40 developers from their trained model. 

%Since in this model, the granularity level of documents in trainset is per commit, for any new author, we can infer vectors for the API content of her/his commits, not its author. 
One solution to predict the embedding vectors for the rest of the developers in our dataset, using this pre-trained model, could be using the embedding vectors of APIs learned in this model. We concatenate the vectors of all APIs in the list of each developer's contributions. Similar to the inference stage of doc2vec, we ignore the API calls which are not seen by this pre-trained model. We calculate the average of all these API embedding vectors to build a single vector that represents the domain expertise of the developer. The bottom pipeline in Figure~\ref{fig:approach} shows the steps of \textit{dev2vec:APIs}.

%There are different advantages in this method. First, we keep the repetition of APIs in developers skill space by merging the content of all commits. Second, we represent same API with same vector in different commits and for different developers because we fetch the learned embedding vectors for each API from the pretrained model. 

%One naive solution could be inferring vector per each commit of a new developer, then concatenate all embedding vectors of commits to build a unique vector that represent the expertise of this developer. We know that the inferring stage in doc2vec can predict different vector for the same document in different attempts~\cite{le2014distributed}. However, the cosine similarity between these vectors is high but they are not the same. We explain with an example how this fact can impact the performance of predicted vectors for a developer. Suppose that two developers submitted 2 different commits on the same source file. None of these commits changed the list of APIs in this source file. However, in inferring stage, the doc2vec model predicts 2 different embedding vectors for these 2 commits with the same content. This small differences between embedding vectors that obtained from same content can add noises into the final vector that represent developers expertise.

\subsection{Dev2vec:RIAs}\label{sec:concat}

\begin{figure*}
  \includegraphics[scale=0.5]{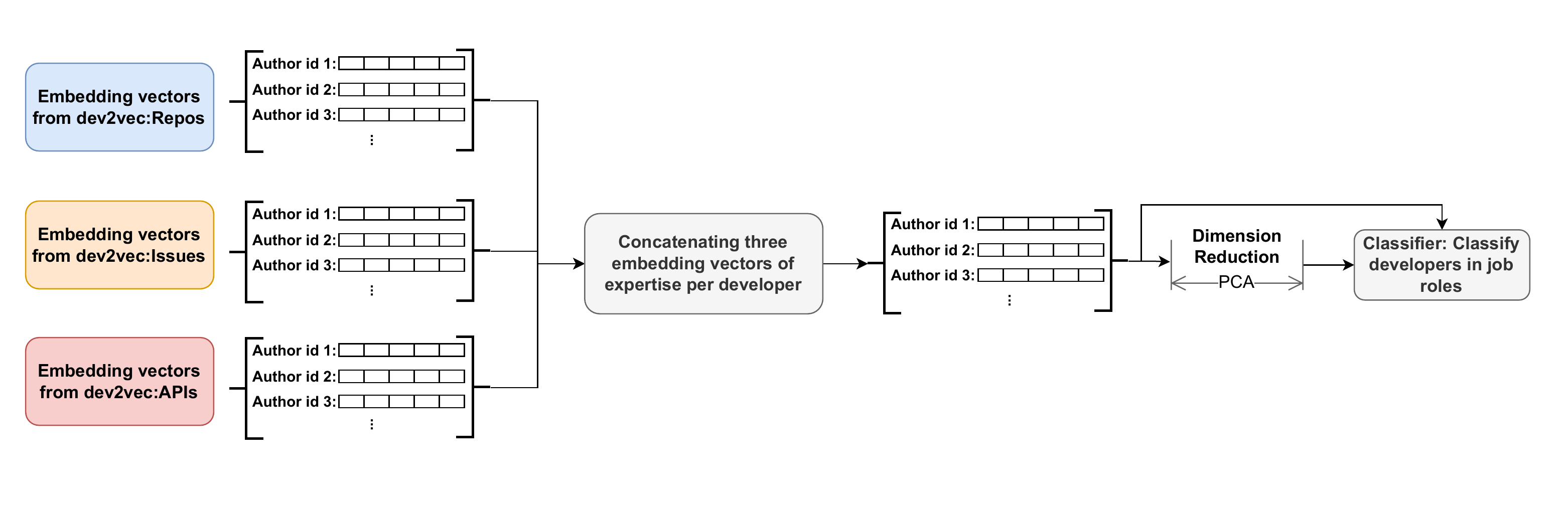}
  \caption{\textbf{Dev2vec:RIAs.} We concatenate three embedding vectors generated in three different spaces of \textit{dev2vec:Repos}, \textit{dev2vec:Issues} and \textit{dev2vec:APIs} to represent the expertise of developers}
  \label{fig:pca}
\end{figure*}

The previous sections generate embedding vectors from single sources.  We now turn to a method to aggregate sources.  Two of the data sources are in natural language and one is in the programming language. We merge sources by concatenating the embedding vectors of developers expertise which are generated from these three spaces. Figure~\ref{fig:pca} shows the pipeline for this section. 

Suppose that the embedding size of vectors generated by \textit{dev2vec:Repos}, \textit{dev2vec:Issues} and \textit{dev2vec:APIs} are ``$s_{repo}$'', ``$s_{issue}$'' and ``$s_{api}$'' respectively. Then, the size of the concatenated vector is ``$s_{repo} + s_{issue} + s_{api}$''. In Section~\ref{sec:setup}, we return to the size of embedding vectors.

Additionally, after concatenation, we explore dimension reduction to see if we can obtain performance improvements. We apply PCA for this purpose. Then, we use both concatenated vectors before and after dimension reduction and compare their performance in representing developers' expertise.

%%%%%%%%%%%%%%%%%%%%%%%%%%%%%%%%%%%%%%%%

\section{Evaluation}\label{eval}
In this section, we first describe how we evaluate the efficiency of each dev2vec by the technical specialization of developers. Then we describe a state-of-the-art method as the  baseline comparison method and explain the setup for each dev2vec model. Finally, we discuss our research questions in light of the results.

\subsection{Technical Specialization of Developers}
To evaluate the efficiency of each dev2vec model to represent the expertise of developers, we choose a common task in software engineering: the classification of developers into job roles.  We use five roles: \textit{Backend, Frontend, Mobile, DevOps, and Data Scientist}.  As shown in Figures~\ref{fig:approach} and~\ref{fig:pca}, we feed embedding vectors derived from each of the models to classifiers to learn the developer role. 

We use three different and well-known classifiers: SVM, Random Forest, and Logistic Regression. We compare the performance of these models by measuring the \textit{Precision, Recall} and \textit{F1-score}.

\subsection{Comparable Method}\label{bs}
Numerous studies used the bag-of-words technique to represent the expertise of developers. We compare our proposed methods to the sate-of-the-arts techniques that focuse more specifically on domain expertise of developers across different projects on GitHub. We replicate a recent study~\cite{montandon2021mining} that focuses on representing the expertise of developers using the information on GitHub. They worked on solving the classification problem of developers in their job roles. Also, we use the same list of developers in their dataset who are categorized into five job roles to train dev2vecs.  %\Foutse{the following sentence is unclear...which study are you referring too..}
%They used the same study that we refer to their labeled developers.

Motandon et al.~ \cite{montandon2021mining} used GitHub information of developer contributions across different projects to represent their expertise. They collected the biography of developers, repositories names, the programming languages of repositories, repositories topics, and repositories dependencies (libraries) across different projects as a group of information that evident the domain expertise of developers. They used the bag-of-words technique to build a vector for each developer based on the frequency of each feature in their contributions. Because of the high dimension of the collected data, they used feature selection with a correlation technique over features from each category (repository names, topics, languages, and dependencies). They omitted features with a high degree of correlation in each category. They end up with 1471 features out of almost 18K. We also applied tf-idf to weight the frequency of these features and call it ``SOA:bow''.

\subsection{dev2vec Setup}\label{sec:setup}

Doc2vec has several hyper-parameters. We empirically optimize important parameters and keep the default values for the rest. The first parameter is the vector size or embedding size. The words and documents embedding vector share the same size. Another parameter is window size which defines the number of words considered around a target word to learn its embedding. Another important parameter is the minimum frequency for a word to be considered (min\_count). Another parameter is negative sampling size. The window size around a word considers positive sampling words for the target word in doc2vec. The words out of this window are considered as negative labels or uncorrelated words to the target words. Since the size of vocabularies is enormous, the negative sampling technique randomly downsampled these words into negative sampling size to speed up the convergence. Finally, we should find the number of epochs to repeat the training and the inference stages. For \textit{dev2vec:APIs} model, we use the pre-trained model in~\cite{dey2021representation}. This model uses DBOW to %\Foutse{preserve the orders...}
do not attempt to the order of APIs because the order of APIs in a source file is not important. %\Foutse{this is strange...if order is not important, why preserve it?}. 

%For \textit{dev2vec:Repos}  and  \textit{dev2vec:Issues}, we empirically cross validate the parameters using the validation set. Table~\ref{tab:setup} shows the final setup for each model. We use DM algorithm for both \textit{dev2vec:Repos} and \textit{dev2vec:Issues} because the order of words is important in such context. 

For cross-validation, we divide our developers into different sets: 80\% of the trainset, 10\% of the validation set, and 10\% of the testset. Developers in testset are not in trainset or validation set. To cross-validate the result, for \textit{dev2vec:Repos} and \textit{dev2vec:Issues}, we search for the best parameter through cross validation over the validation set. After learning the vector representations for the developers in trainset, we feed them to a classifier to learn a predictor of the developers job roles. For the inference stage, we infer vectors for developers in testset through these two dev2vec models and use them as separate testsets of the classifiers. 

For \textit{dev2vec:API}, since there is no inference stage, we build the embedding vectors of the developers by averaging the embedding vector of APIs in the list of their API expertise. We use the embedding vectors of developers in trainset to train the classifier and the embedding vector of developers in testset to test it. We apply the same process for the comparable method, ``SOA:bow''.

\begin{table}[htbp]
   \centering
   \noindent
  \caption{Experimental Setup for dev2vec models}
\resizebox{.99\textwidth}{!}{\begin{tabular}{cccccccc} 
\toprule
\textbf{Model}& \textbf{vector size}&	\textbf{window size}&	\textbf{min\_count}&	\textbf{algorithm}&	\textbf{negative sampling}&	\textbf{epochs}

\\
\midrule
\textbf{dec2ve:Repos}&	230&	5&	5&	DM &	5(default)&	15
 \\
 \midrule
\textbf{dev2vec:Issues}	& 150 &	5 &	5&	DM &	5(default)&	20
\\
 \midrule
 \textbf{dev2vec:APIs}&	200&	30&	5&	DBOW&	20&	10(defualt)
\\
\bottomrule
\end{tabular}
}
  \label{tab:setup}
\end{table}

%%%%%%%%%%%%%%%%%%%%%%%%%%%%%%%%%%%%%%%%%%%%%
\subsection{Experimental Result}
In this section, we address our research questions by comparing dev2vec methods with a recent state-of-the-art method, investigating three different dev2vec models learned from different sources of information, and the combination of embedding vectors learned from three different spaces.

\subsubsection{RQ1: How effective are embedding vectors to represent the domain expertise of developers across various software projects compared to the state-of-the-art methods?}

To answer RQ1, we consider the method described in Section~\ref{bs}, as a state-of-the-art method, ``SOA:bow'' for comparison purpose.  Since the distribution of developers in different job roles is imbalanced and 41\% of developers are \textit{``frontend''}, the baseline is a majority-rule classifier that predicts all developers as \textit{``frontend''}. For SOA:bow and all dev2vecs, we apply Random Forest to classify developers in different job roles. 

Since the classes in our dataset are imbalance, we apply Macro-Weighted Precision, Recall and F1-score to report the performance. With Macro-Weighted metrics, the proportion of each class is weighted by its relative number of examples available in dataset. 

\begin{table}[htbp]
   \centering
   \noindent
  \caption{The performance of a classifier across five job roles based on three dev2vec methods, \textit{dev2vec:Repos}, \textit{dev2vec:Issues} and \textit{dev2vec:APIs}, compare to a sate-of-the-art, \textit{SOA:bow}, and a baseline. Since the classes are imbalance, Macro-Weighted Precision, Recall and F1-score are used to report the performance. }
\resizebox{.99\textwidth}{!}{\begin{tabular}{cccc} 
\toprule
\textbf{Methods}& 	\textbf{Macro-Weighted Precision\%}&	\textbf{Macro-Weighted Recall\%}&	\textbf{Macro-Weighted F1-score \%}

\\
\midrule
\textbf{Baseline}&	18.73&	43.28&	26.15 
 \\
\midrule
\textbf{SOA:bow}&	41.73&	43.93&	42.35
 \\
\midrule
\textbf{dev2vec:Repos}&	60.78&	61.90&	\textbf{60.02}
 \\
 \midrule
\textbf{dev2vec:Issues}	& 	64.75 &	65.04&	\textbf{63.08}
\\
 \midrule
 \textbf{dev2vec:APIs}&	56.90&	55.10&	\textbf{55.51}
\\
\bottomrule
\end{tabular}
}
  \label{tab:macro_result}
\end{table}

We can find in Table~\ref{tab:macro_result} that all dev2vec methods and the state-of-the-art, SOA:bow, show a better performance than the baseline. The Macro-Weighted Precision to classify developers in their job roles with SOA:bow is 41.73\%, while it is equal to 18.73\% for the baseline. This precision is equal to 60.78\%, 64.75\% and 56.90\% for \textit{dev2vec:Repos}, \textit{dev2vec:Issues} and \textit{dev2vec:APIs}, respectively.

\textit{dev2vec:Issues} shows 64.75\% Macro-Weighted Precision for classifying developers in their job roles. It improves the precision of the classification task up to 23.02\% compared to the SOA:bow. The Macro-Weighted F1-score for \textit{dev2vec:Repos}, \textit{dev2vec:Issues} and \textit{dev2vec:APIs} equals 60.02\%, 63.08\%, 55.51\% respectively compare to the SOA:bow with 42.35\%  F1-score.  %\Foutse{too many 'to'...} 

\begin{tcolorbox}[width=1\textwidth]
\textit{Answer to RQ1:} Representing the domain expertise of developers across different software projects in embedding vectors shows better performance than both a baseline and a recent state-of-the-art in classifying developers in their technical job roles. The proposed dev2vec methods improve F1-Score at least 13.16\% with \textit{dev2vec:APIs} and at most 20.73\% with \textit{dev2vec:Issues}.
\end{tcolorbox}

%%%%%%%%%%%%%%%%%%%%%%%%%%%%%%%%%%
\subsubsection{RQ2: How sensitive is the performance of dev2vec to the source of information (repositories, issues and API)?}\label{sec:RQ2}

To answer RQ2, we compare three different dev2vec models, \textit{dev2vec:Repos}, \textit{dev2vec:Issues},  and \textit{dev2vec:APIs}, that are trained on three different sources of activities to investigate their impact on representing the expertise of developers in embedding vectors.

Table~\ref{tab:final_result} shows the result of all three dev2vec methods for classifying developers with three types of classifiers. We apply SVM, Random Forest and Logistic Regression. The metrics are measured in each job role to study the difference in performance between different classes separately. The table also presents the results of SOA:bow on each job role.

Among the three dev2vec models, \textit{dev2vec:Issue} shows better performance than \textit{dev2vec:Repos} and \textit{dev2vc:APIs} in classifying developers in different job roles. For example, the precision of \textit{dev2vec:Issue} in classifying developers with \textit{Data Scientist} role is 80\% and the recall for \textit{Frontend} developers is 90.48\%.  It also improves the F1-score for  \textit{DevOps} developers up to 26.82\% with Logistic Regression classifier. The \textit{dev2vec:repos} shows better performance than \textit{dev2vc:APIs}. For example, \textit{dev2vec:APIs} has 46.43\% and 36.11\% precision and recall respectively for \textit{Mobile} developers with SVM. However, it improves to 53.82\% and 46.42\% with \textit{dev2vec:Repos}. 

In addition to the above comparison, all three dev2vec models improve the performance of the classification task in all job roles compared to the state-of-the-art, \textit{SOA:bow}. For example, in \textit{SOA:bow} and SVM classifier, recall for \textit{Backend} and \textit{DevOps} developers are 20\% and 23.08\%, respectively. It increases into 31.25\% and 44.45\% with dev2vec models. As another example, the precision for \textit{Data Scientist} class in bag-of-words with Random Forest classifier is 46.15\% and it increases into 78.57\%, 80\% and 64.71\% with \textit{dev2vec:Repos}, \textit{dev2vec:Issues}, and \textit{dev2vec:APIs}, respectively.

% We can sort the dev2vec methods based on their performance into the order of \textit{dev2vec:Issues}, \textit{dev2vec:Repos}, and \textit{dev2vec:APIs}. The explanation for these results is that the description of repositories maybe is shared between developers with different job roles involved in the same project, but the issue resolving history shows more details about the contribution of developers. In \textit{dev2vc:APIs}, we are not collecting the APIs that developers practiced in the past.  Also, representing the expertise of developers by averaging API embedding vectors into one vector could impact the performance of the final vector.

Some job roles show better performance. For example, ``Data Scientist'' developers are more accurately classified compared to ``Backend'' developers. This is due to the intersection between the activities of developers in different roles. We discuss it more in Section~\ref{sec:TV}. 

\begin{table}[htbp]
\centering
\noindent
\vspace{-10pt}
\caption{The result of three different classifiers on embedding vector representation of developers expertise in three different spaces. The ``Pre'', ``Rec'' and ``F1'' are referring to Precision, Recall and F1-score, respectively.}
\resizebox{.99\textwidth}{!}{
\begin{tabular}{N N N N N N N N N N N N N}
\toprule
\multicolumn{1}{N }{\textbf{}} &\multicolumn{12}{c }{\textbf{SVM}}\\
\cmidrule(lr){2-13}
\multicolumn{1}{N }{\textbf{}} &\multicolumn{3}{c }{\textbf{SOA:bow}}& \multicolumn{3}{c }{\textbf{Dev2vec:Repos}} & \multicolumn{3}{c }{\textbf{Dev2vec: Issues}} & \multicolumn{3}{c }{\textbf{Dev2vec: APIs}} \\
\cmidrule(lr){1-1}
\cmidrule(lr){2-4}
\cmidrule(lr){5-7}
\cmidrule(lr){8-10}
\cmidrule(ll){11-13}
\multicolumn{1}{ c }{\textbf{Job Role}}
      &  \shortstack[c]{\textbf{Prec\%}} &
      \shortstack[c]{\textbf{Rec\%}} &
      \shortstack[c]{\textbf{F1\%}} &  
      \shortstack[c]{\textbf{Prec\%}} &  \shortstack[c]{\textbf{Rec\%}} &   \shortstack[c]{\textbf{F1\%}} & 
      \shortstack[c]{\textbf{Prec\%}} &
      \shortstack[c]{\textbf{Rec\%}} &  \shortstack[c]{\textbf{F1\%}}& 
      \shortstack[c]{\textbf{Prec\%}} &
      \shortstack[c]{\textbf{Rec\%}} &  \shortstack[c]{\textbf{F1\%}} \\  
      
\cmidrule(lr){1-1}
\cmidrule(lr){2-4}
\cmidrule(ll){5-7}
\cmidrule(ll){8-10}
\cmidrule(ll){11-13}

\multicolumn{1}{ l }{\textbf{Frontend}} & 53.52&	73.08&	61.79&	63.28&	85.40&	72.69&	\textbf{66.67}&	\textbf{90.48}&	\textbf{76.77}&	61.11&	78.57&	68.75
 \\
\multicolumn{1}{ l }{\textbf{Backend}} &      33.33&	20.00&	25.00&	37.50&	28.64&	31.79&	\textbf{55.56}&	\textbf{31.25}&	\textbf{40.00}&	31.25&	27.78&	29.41
 \\
 \multicolumn{1}{ l }{\textbf{Mobile}} & 32.26&	30.30&	31.25&	53.82&	46.42&	48.17&	\textbf{59.09}&	\textbf{52.00}&	\textbf{55.32}&	46.43&	36.11&	40.63
 \\
\multicolumn{1}{ l }{\textbf{DataScientist}} &     60.00&	42.86&	50.00&	76.04&	61.76&	67.82&	\textbf{80.00}&	\textbf{72.73}&	\textbf{76.19}&	66.67&	66.67&	66.67
 \\
\multicolumn{1}{ l }{ \textbf{DevOps}} &      37.50&	23.08&	28.57&	68.89&	37.17&	48.08&	\textbf{80.00}&	\textbf{44.45}&	\textbf{57.14}&	37.50&	30.00&	33.33
 \\
 \cmidrule(ll){1-13}
\multicolumn{1}{N }{\textbf{}} &\multicolumn{12}{c }{\textbf{Random Forest}}\\
\cmidrule(ll){2-13}
\multicolumn{1}{N }{\textbf{}} &\multicolumn{3}{c }{\textbf{SOA:bow}}& \multicolumn{3}{c }{\textbf{Dev2vec:Repos}} & \multicolumn{3}{c }{\textbf{Dev2vec: Issues}} & \multicolumn{3}{c }{\textbf{Dev2vec: APIs}} \\

\cmidrule(lr){1-1}
\cmidrule(lr){2-4}
\cmidrule(lr){5-7}
\cmidrule(lr){8-10}
\cmidrule(ll){11-13}

\multicolumn{1}{ c }{\textbf{Job Role}}
      &  \shortstack[c]{\textbf{Prec\%}} &
      \shortstack[c]{\textbf{Rec\%}} &
      \shortstack[c]{\textbf{F1\%}} &  
      \shortstack[c]{\textbf{Prec\%}} &  \shortstack[c]{\textbf{Rec\%}} &   \shortstack[c]{\textbf{F1\%}} & 
      \shortstack[c]{\textbf{Prec\%}} &
      \shortstack[c]{\textbf{Rec\%}} &  \shortstack[c]{\textbf{F1\%}}& 
      \shortstack[c]{\textbf{Prec\%}} &
      \shortstack[c]{\textbf{Rec\%}} &  \shortstack[c]{\textbf{F1\%}} \\  
      
\cmidrule(lr){1-1}
\cmidrule(lr){2-4}
\cmidrule(ll){5-7}
\cmidrule(ll){8-10}
\cmidrule(ll){11-13}

\multicolumn{1}{ l }{\textbf{Frontend}} &  55.56&	68.18&	61.22&	64.29&	84.91&	73.17&	\textbf{67.27}&	\textbf{88.10}&	\textbf{76.29}&	56.00&	73.68&	63.64
 \\
\multicolumn{1}{ l }{\textbf{Backend}} &  31.25&	20.00&	24.39&	50.00&	27.27&	35.29&	\textbf{55.56}&	\textbf{31.25}&	\textbf{40.00}&	33.33&	31.25&	32.26
 \\
 \multicolumn{1}{ l }{\textbf{Mobile}} & 38.46&	40.54&	39.47&	53.13&	45.95&	49.28&	\textbf{56.00}&	\textbf{56.00}&	\textbf{56.00}&	50.00&	41.03&	45.07
 \\
\multicolumn{1}{ l }{\textbf{DataScientist}} &  46.15&	46.15&	46.15&	78.57&	64.71&	70.97&	\textbf{80.00}&	\textbf{72.73}&	\textbf{76.19}&	64.71&	64.71&	64.71
 \\
\multicolumn{1}{ l }{ \textbf{DevOps}} & 20.00&	15.38&	17.39&	50.00&	25.00&	33.33&	\textbf{75.00}&	\textbf{33.33}&	\textbf{46.15}&	30.00&	21.43&	25.00
 \\

  \cmidrule(ll){1-13}
\multicolumn{1}{N }{\textbf{}} &\multicolumn{12}{c }{\textbf{Logistic Regression}}\\
\cmidrule(ll){2-13}
\multicolumn{1}{N }{\textbf{}} &\multicolumn{3}{c }{\textbf{SOA:bow}}& \multicolumn{3}{c }{\textbf{Dev2vec:Repos}} & \multicolumn{3}{c }{\textbf{Dev2vec: Issues}} & \multicolumn{3}{c }{\textbf{Dev2vec: APIs}} \\
\cmidrule(lr){1-1}
\cmidrule(lr){2-4}
\cmidrule(lr){5-7}
\cmidrule(lr){8-10}
\cmidrule(ll){11-13}

\multicolumn{1}{c}{\textbf{Job Role}}
      &  \shortstack[c]{\textbf{Prec\%}} &
      \shortstack[c]{\textbf{Rec\%}} &
      \shortstack[c]{\textbf{F1\%}} &  
      \shortstack[c]{\textbf{Prec\%}} &  \shortstack[c]{\textbf{Rec\%}} &   \shortstack[c]{\textbf{F1\%}} & 
      \shortstack[c]{\textbf{Prec\%}} &
      \shortstack[c]{\textbf{Rec\%}} &  \shortstack[c]{\textbf{F1\%}}& 
      \shortstack[c]{\textbf{Prec\%}} &
      \shortstack[c]{\textbf{Rec\%}} &  \shortstack[c]{\textbf{F1\%}} \\  
    
\cmidrule(lr){1-1}
\cmidrule(lr){2-4}
\cmidrule(ll){5-7}
\cmidrule(ll){8-10}
\cmidrule(ll){11-13}

\multicolumn{1}{ l }{\textbf{Frontend}} &  46.43&	47.27&	46.85&	74.47&	68.63&	71.43&	\textbf{76.92}&	\textbf{71.43}&	\textbf{74.07}&	60.53&	50.00&	54.76
\\
\multicolumn{1}{ l }{\textbf{Backend}} &  23.53&	19.05&	21.05&	25.00&	30.77&	27.59&	\textbf{50.00}&	\textbf{31.25}&	\textbf{38.46}&	20.00&	28.57&	23.53
 \\
 \multicolumn{1}{ l }{\textbf{Mobile}} & 37.50&	40.00&	38.71&	50.00&	43.24&	46.38&	\textbf{53.13}&	\textbf{68.00}&	\textbf{59.65}&	42.86&	42.86&	42.86
 \\
\multicolumn{1}{ l }{\textbf{DataScientist}} &  40.00&	46.15&	42.86&	62.50&	58.82&	60.61&	\textbf{66.67}&	\textbf{72.73}&	\textbf{69.57}&	50.00&	52.63&	51.28
 \\
\multicolumn{1}{ l }{ \textbf{DevOps}} &	25.00&	23.08&	24.00&	26.67&	50.00&	34.78&	\textbf{50.00}&	\textbf{55.56}&	\textbf{52.63}&	25.00&	26.67&	25.81
 \\
\bottomrule
\end{tabular}
}
\label{tab:final_result}
\end{table}

\begin{tcolorbox}\textit{Answer to RQ2:} The performance of dev2vec model is sensitive to embedding vectors learned from different sources of developers' activities. The \textit{dev2vec:Issues} shows a better performance in representing the expertise of developers in embedding vectors. The \textit{dev2vec:Repos} and \textit{dev2vec:APIs} are in second and third place, respectively.
\end{tcolorbox}

%\vspace{-10pt}

%%%%%%%%%%%%%%%%%%%%%%%%%

\subsubsection{RQ3: How effective is the concatenation of expertise embedding vectors from the different information sources?}

We concatenate three embedding vectors from three different spaces of their activities, as explained in Section~\ref{sec:concat}. The size of the final embedding vector that represents the expertise of developers is~$580$. 

We feed these vectors into a Random Forest classifier to predict developer job roles. We also apply PCA on these embedding vectors to reduce their dimensionality from $580$ to $50$, $100$, $200$, $250$, and $300$, and each time, train a new classifier for predicting the developers job roles. Table~\ref{tab:PCA_result} shows the results for embedding vectors with and without dimension reduction. Based on this table, concatenating embedding vectors from different spaces improves the performance of the classifier only in two job roles, ``Frontend'' and ``Data Scientist''. But for the rest of job roles the performance is as good as \textit{dev2vec:Issues}. Applying PCA to reduce dimension has an adverse effect on the performance of the classifier. Increasing the number of dimensions improves the performance. However, even with ``PCA-300'', the performance of classifier is not as good as \textit{dev2vec:Issues}.

\begin{tcolorbox}\textit{Answer to RQ3:} Concatenating embedding vectors from different spaces to represent the expertise of developers, without dimension reduction, improves the performance of the classifier for two job roles, and for the rest of the roles, the performance is as good as \textit{dev2vec:Issues}.
\end{tcolorbox}

\begin{table}[htbp]
\centering
\noindent
%\vspace{-10pt}
\caption{The performance of \textit{dev2vec:RIAs} with different dimensionality reduction levels in classifying developers in their job roles}
%\resizebox{\textwidth}{!}{
\vspace*{1ex}
\resizebox{.5\textwidth}{!}{
  \begin{tabular}{clccc}
\toprule

%\multicolumn{1}{ c }
{\textbf{Reduction}} & {\textbf{Roles}}
&{\textbf{Prec\%}}&{\textbf{Rec\%}}&{\textbf{F1\%}}\\

\cmidrule(lr){1-5}

\multirow{4}{*}{\textbf{None}} &{\textbf{Frontend}} & 69.09 &	88.37&	\textbf{77.55}
 \\
 & \textbf{Backend} & 55.56&	31.25&	40.00
 \\
&{\textbf{Mobile}} & 56.00&	56.00&	56.00
 \\
&{\textbf{DataScientist}} & 81.82&	81.82&	\textbf{81.82}
 \\
 
&{\textbf{DevOps}} & 75.00&	33.33&	46.15
 \\
\cmidrule(lr){1-5}

\cmidrule(lr){1-5}

\multirow{4}{*}{\textbf{PCA-50}} &{\textbf{Frontend}} & 55.36&	73.81&	63.27
 \\
 & \textbf{Backend} & 30.77&	23.53&	26.67
 \\
&{\textbf{Mobile}} & 40.00&	32.00&	35.56
 \\
&{\textbf{DataScientist}} & 58.33&	58.33&	58.33
 \\
 
&{\textbf{DevOps}} & 40.00 &20.00&	26.67
\\

 \cmidrule(lr){1-5}

\cmidrule(lr){1-5}

\multirow{4}{*}{\textbf{PCA-100}} &{\textbf{Frontend}} & 56.14&	76.19&	64.65
 \\
 & \textbf{Backend} & 30.77&	23.53&	26.67
 \\
&{\textbf{Mobile}} & 42.11 &32.00&	36.36
 \\
&{\textbf{DataScientist}} & 58.33&	58.33&	58.33
 \\
 
&{\textbf{DevOps}} & 40.00 &20.00&	26.67
\\
 
 \cmidrule(lr){1-5}

\cmidrule(lr){1-5}

\multirow{4}{*}{\textbf{PCA-200}} &{\textbf{Frontend}} & 60.71&	80.95&	69.39
 \\
 & \textbf{Backend} & 38.46&	29.41&	33.33
 \\
&{\textbf{Mobile}} & 47.37&	36.00&	40.91
 \\
&{\textbf{DataScientist}} & 58.33&	58.33&	58.33
 \\
 
&{\textbf{DevOps}} & 50.00 & 30.00&	37.50
\\

 \cmidrule(lr){1-5}

\cmidrule(lr){1-5}

\multirow{4}{*}{\textbf{PCA-250}} &{\textbf{Frontend}} & 61.82&	80.95&	70.10
 \\
 & \textbf{Backend} & 38.46&	29.41&	33.33
 \\
&{\textbf{Mobile}} & 47.37 & 36.00&	40.91
 \\
&{\textbf{DataScientist}} &61.54&	66.67&	64.00
 \\
 
&{\textbf{DevOps}} & 50.00&	30.00&	37.50
\\

\cmidrule(lr){1-5}

\cmidrule(lr){1-5}

\multirow{4}{*}{\textbf{PCA-300}} &{\textbf{Frontend}} & 62.50&	83.33&	71.43
 \\
 & \textbf{Backend} & 41.67&	29.41&	34.48
 \\
&{\textbf{Mobile}} & 50.00 &40.00&	44.44
 \\
&{\textbf{DataScientist}} &61.54&	66.67&	64.00
 \\
 
&{\textbf{DevOps}} & 60.00& 30.00&	40.00
\\

\bottomrule
\end{tabular}}
%}
\label{tab:PCA_result}
\end{table}

%%%%%%%%%%%%%%
\section{Threat to Validity}\label{sec:TV}

In this section, we explain the scopes and assumptions in our study that can threaten the validity of our results.

\subsection{Data Assumption}
For \textit{dev2vec:APIs}, we use a pre-trained doc2vec model on API list collected from source files that were modified by developers with different commits. Thus, we use the same approach as~\cite{dey2021representation} to collect the API calls from developers' commits. We assume that if developers submit a commit on a file, they have a basic knowledge about the libraries used in that source file~\cite{dey2021representation}. With this assumption, we are not collecting the actual practice of developers. Because the developer may not change the list of APIs in a source file by submitting a commit. Therefore, the performance of embedding vectors derived from \textit{dev2vec:APIs} to represent the expertise of developers can be affected by this assumption.

Further along, in \textit{dev2vec:repos}, sometimes, developers with different domains of expertise contribute to the same repositories. Thus, we may collect the same content (the textual information of repositories) for developers with different domains who contributed to the same project. This assumption can impact the performance of embedding vectors derived from \textit{dev2vec:repos}.

\subsection{Averaging API embedding vectors}
In \textit{dev2vec:APIs}, we cannot directly use the pre-trained model in~\cite{dey2021representation} to predict the embedding vectors of expertise for new developers. Thus, we average the embedding vectors of APIs in the list of APIs related to a developer's activities and represent the expertise of the developer. 

One naive solution could be inferring a vector per each commit of a new developer. Then, merge all embedding vectors of commits to build a unique vector that represents the expertise of the developer. But, we know that the inference stage in doc2vec can predict different vectors for the same document in different attempts~\cite{le2014distributed}. However, the cosine similarity between these vectors is high, but they are not the same. We explain with an example how this fact can impact the performance of predicting vectors for a developer. Suppose that two developers submitted two different commits on the same source file. None of these commits changed the list of APIs in this source file. However, in the inference stage, the doc2vec model predicts two different embedding vectors for these two commits with the same content. This small difference between embedding vectors (that are obtained from the same content) can add noise into the final vectors of expertise.

There are advantages to our proposed method, \textit{dev2vec:APIs}. First, we keep the repetition of APIs in developers' activities by merging the API calls in the source files of all commits. Second, we represent the same API with the same vector in different commits and for different developers because we fetch the learned embedding vectors for each API from the pre-trained model. We can consider this averaging as a weighted average. It means if a developer uses two APIs more than the rest of the APIs in her activities, the final embedding vector would be more similar to the average of embedding vectors of these two APIs. However, this averaging can decrease the impacts of other APIs (rare ones) in defining the expertise of developers. In future studies, we can use other mechanisms similar to the attention mechanism to learn weights for aggregating the API vectors that can yield a more accurate vector to represent developer expertise.

\subsection{Type and source of activities} 

In this study, we focus on three well-known types of developers' activities on GitHub. We perform that the performance of representing the domain expertise of developers in embedding vectors is sensitive to the source of activities. In addition to activities that we considered in this work, there are other activities on GitHub that reflect developers' expertise, such as commit messages submitted by developers, the list of APIs that developers practiced in a commit, or the structure of codes written by them. Also, GitHub is not the only source of developers' activities. Other platforms such as StackOverflow contain valuable information about the domain expertise of developers as well. Other activities on GitHub and other platforms such as StackOverflow can be investigated in future studies to learn embedding vectors of expertise.

%In this study, we learn that the performance of embedding vectors in representing the domain expertise of developers is sensitive to the input data fed into the doc2vec models. Our result shows embedding vectors of expertise that derived from issue resolving history of developers has a better performance in reflecting their expertise. To extend this study, we investigate how the combination of these three types of embedding vectors, dev2vec:Repos, dev2vec:Issues and dev2vec:APIs impacts the performance of representing developers expertise. 

%\Foutse{any lesson learned from experimenting with these embeddings? anything on the modeling?}

\subsection{Job roles of developers} 

The pre-labeled dataset that we used in this study categories a group of developers in five job roles.  However, the job roles of developers are not limited to these five roles, but the investigation of authors in~\cite{montandon2021mining} showed that 64\% of job posts on StackOverflow are related to one of these five roles.

\section{Discussion}\label{disc}
%The expertise of developers has different aspects from soft skills such as teamwork or communication to technical skills such as programming languages, algorithm design, libraries, etc. In this study, we focus on the technical expertise of developers. Categorizing developers in certain domains is a reasonable initial filtering for recruiters or project managers. In addition to this initial categorization, other investigation such as interview is required to choose a proper candidate. 

In this section first we discuss why we choose the combination of doc2vec to represent the expertise of developers and machine learning algorithms for classification task. Then, we discuss why the performance of dev2vec models varies across different job roles.

\subsection{Combination of doc2vec and machine learning classifiers} 

This study sheds light on representing the domain expertise of developers across different projects with embedding vectors. We apply doc2vec to developers and learn the embedding vector representation of their expertise. Then, we use this model to predict embedding vectors for those developers. In the last step, we investigate the performance of these vectors in representing developers' expertise with the Random Forest classifier.  We exclude an end-to-end neural network for the classification task because the size of the dataset is not suitable to train or train a transformer for representing the expertise of developers. 

\subsection{The variety of the performance of the models in different job roles} 

As Table~\ref{tab:final_result} results show, the performance of classifiers in job roles such as Frontend and Data Scientist is better than Backend or Mobile for \textit{SOA:bow} and all dev2vec models.  In practice, we know that there are intersections between the skills of developers. For example, there are confluences between the skills and projects of Frontend developers and Backend developers or Mobile developers. However, this confluence is smaller when it comes to Data Scientists compared to Frontend or Mobile developers.

To illustrate the correlation between different job roles, we investigate the similarity between words across topics with the \textit{Cosine Inter\_Intra Topics} method~\cite{neishabouri2021estimating}. We apply this method to the state-of-the-art \textit{SOA:bow} dataset, which we adopted as our ground truth.  As discussed in Section~\ref{bs}, their dataset includes words collected from the biography of developers and repository names, languages, topics, and dependencies (libraries) across different projects on GitHub after  removing highly correlated words.

\begin{figure}
\centering
  \includegraphics[width=0.7\textwidth]{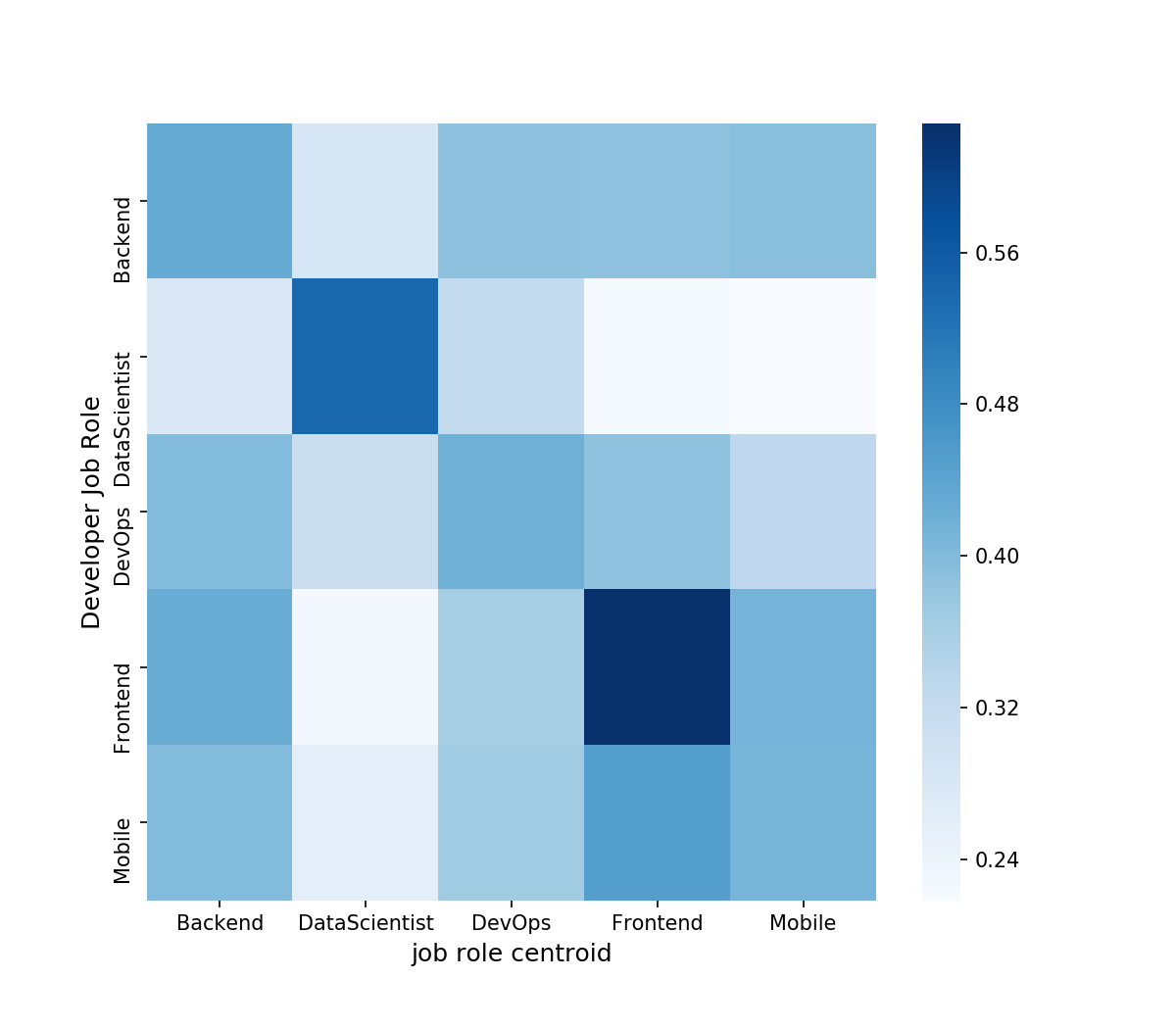}
  \caption{\textbf{Cosine Inter\_Intra Topics (job roles).} Rows are the roles and columns are the role's centroid. In ``Frontend'' or ``Data Scientist'' roles, the similarity between developers within the role is more significant than developers across different roles. However, the inter similarity for job roles such as ``Mobile'' or ``Backend'' is very close to their intra similarity with others. It impacts the performance of the classifier for these roles.}
  \label{fig:corr}
\end{figure}

In Figure~\ref{fig:corr}, rows represent the job roles, and columns represent the centroid of each role.
We  find that the diagonal values for Frontend and Data Scientist roles are greater than off-diagonal values. This implies that the similarity for developers within these two job roles is greater than the similarity between them and developers across other roles. However, for Backend and Mobile developers, the diagonal and off-diagonal values are very close. Even this similarity between Mobile and Frontend developers is more significant than the similarity among Mobile developers. It could explain the lower performance of classifiers in Backend and Mobile roles. 

\subsection{Implication}
Assessing the expertise of developers has different aspects. This assessment includes their soft skills, such as teamwork or communication, and technical skills, such as programming languages, algorithm design, and libraries. In this study, we focus on representing the domain expertise of developers across different projects. This study is initial filtering for recruiters or project managers. In addition to this initial categorization, more investigations, such as interviews, are required to choose a proper candidate.

Moreover, we evaluate the performance of embedding vectors of expertise in classifying developers in different job roles, but their benefits are not limited to this task. The embedding vectors of developers' expertise can be applied in various manners in software engineering. For example, we can use these embedding vectors of expertise to find a proper candidate for a job post or a new project contributor. We also can use these embedding vectors to match developers with similar expertise (working in the same domain). These significances can be addressed in future works.

For the sake of replication, all trained models and scripts are made available\footnote{\url{https://github.com/ExpertiseModel/EmbeddingVectors}}.
%We can conclude that to evaluate the performance of different models in such problems, the matter of ground truth is also important and can impact the results.

%\Foutse{Add a lesson learned and discussion section, to explain the meaning of your results, with respect to the problem at hand, embedding, discussing what could be done address the shortcomings, other techniques that could be explored, other problems for which your work can be relevant, etc .....}

%%%%%%%%%%%%%%%%%%%%%%%%%%%
\section{Related Works} \label{sec:RW}
In this section, we review the previous works. First, we discuss the studies focusing on developers expertise. Second, we describe the efficacy of embedding methods in software engineering literature. 

\underline{\textit{Developers Expertise:}} The majority of studies to assess the expertise of developers consider the number of changes or commits in a file path. A widely used heuristic is the \textit{Line 10 Rule}~\cite{mcdonald2000expertise,mockus2002expertise}, inspired by version control systems that store the name of the author in line 10 of the commit log. All methods motivated by \textit{Line 10 Rule} heuristic state that if a developer changed a file in the past, she/he should be one of the candidates to solve the tasks related to this file~\cite{minto2007recommending,tian2016learning, kim2013should,anvik2007Implementation}. These methods cannot be expanded across different projects because they are limited to the path of a source file. 

Other studies go further and show information such as commit messages~\cite{matter2009assigning}, bug resolving history~\cite{anvik2007determining}, the description of repositories and content of ReadMe files on GitHub~\cite{greene2016cvexplorer,wan2018scsminer} or the API calls in source files~\cite{sindhgatta2008identifying}, touched by a developer, are good source of information to infer the expertise of developers. Although, these methods have a good performance to define the expertise of developers within a software project. Collecting these types of information across different software projects increases the number of features and in turn, increases the sparsity that affect the performance of these methods. Montandon et al. in~\cite{montandon2021mining} represent the domain expertise of developers by collecting their activities across different projects. They apply bag-of-word technique on data collected from the biography of developers on Github, repository descriptions, programming languages and project dependencies to represent the expertise of developers. To reduce the sparsity, they reduce the dimension of collected data by calculating the correlation between different features and select high correlated ones. 

A recent study represents the expertise of developers by collecting programming syntax patterns of python code written by them across different software projects~\cite{moradi2021assessing}. This model shows good performance in categorizing developers in two levels of expert and novice. They focus on the levels of expertise of developers in python programming language.\\
%It can be concluded that there are challenges to expand previous studies to assess the domain expertise of developers across different software projects.\\

\underline{\textit{Vector Embedding in Software Domain:}} Recently in software engineering, vector embedding methods have been widely used for different purposes. Zhang et al.~\cite{zhang2020ilinker} use average word embedding and document embedding for issue knowledge acquisition. They derive embedding vectors for each issue from the content in the title and body of them. Then, they link an issue to potentially related issues by calculating the similarity between issue vectors. 

Cod2vec~\cite{alon2019code2vec} projects code into embedding vectors by training a model on different paths collected from the AST of methods and then attempt to predict methods name for code snippets in the testset. Another study applies code2vec pre-trained model on different commits to represent them in embedding vectors and then classify commits vectors into security relevant/non-relevant ones~\cite{lozoya2021commit2vec}. Recently, a new study represented API calls in source file of three programming languages (java, java script and python) in an embedding vector. To investigate the performance of these vectors in representing API calls, they consider all APIs in a source file as related APIs. Then, they find that the vectors of these APIs are more similar than the vector of APIs that never occur together in a source file~\cite{theeten2019import2vec}. Dey et al. \cite{dey2021representation} in a recent study collect API calls in commits of developers in different projects in 17 programming language. They obtain embedding vectors for 3 different entities: APIs, projects and developers. They call this information ''API-related Skill Space`` of projects and developers. Then, they find that developers are more likely to use new APIs or join new projects that have similar representations to themselves in the API-related Skill Space. However, they didn't evaluate their model in predicting the embedding vector for a new developer which is not seen during training phase.

In this study, we identify the domain expertise of developers by representing their expertise in embedding vectors. We investigate the performance of using different source of information such as repositories description, issue resolving history of developers and libraries used by developers to obtain these embedding vectors of expertise.
%%%%%%%%%%%%%%%%%%%%%%%%%%%%%%%%%%%
\section{Conclusion}\label{CL}
%\Foutse{please revise this conclusion...the style is not good...just focus on recalling the goal of your paper, your achieve results, and outline some future works!}
The diversity in the technical specialization of developers in software engineering is expanding. Automatic assessment of developers' specialization is consequential for the success of software projects. Appropriate recognition of this specialization requires an accurate and comprehensive representation of the domain expertise of developers. In this study, we investigate the expertise of developers across their contributions to various software projects and multiple programming languages. Previous methods that show promising outcomes in assessing developers' expertise within a software project are not practical across various projects due to the data size. To address this challenge, we proposed the dev2vec approach pivoted on doc2vec and represented the expertise of developers in embedding space. We trained three models on three different sources of information on GitHub that capture the expertise of developers. We refer to them as \textit{dev2vec:Repos}, \textit{dev2vec:Issues} and \textit{dev2vec:APIs}. Furthermore, we merge the output of these three methods by concatenating the embedding vectors from three spaces and call it \textit{dev2vec:RIAs}. To study the effectiveness of our proposed methods in representing the domain expertise of developers, we employed these embedding vectors in a vital problem in the software engineering domain:  classifying developers in their job roles. Our result on this classification task shows improvements on F1-score at least 13.16\% with \textit{dev2vec:APIs} and at most 20.73\% with \textit{dev2vec:Issues} compare to the state-of-the-art methods. In addition, we observed that the performance of embedding vectors in representing the domain expertise of developers is sensitive to the source of information, and, embedding vectors of expertise that are derived from issue resolving history of developers has a better performance in reflecting their expertise. Furthermore, the performance of concatenation of embedding vectors from different spaces, \textit{dev2vec:RIA}, is as good as \textit{dev2vec:Issues}. 

In future works, we aim to apply mechanisms similar to the attention mechanism and learn a weight vector for concatenating the embedding vectors of expertise. We also want to extract more precisely the content of changes made by the developers to derive these embedding vectors of expertise such as the list of APIs that they practiced or the structure of codes written by them.  Information available on other platforms is also considered. Finally, we plan to study the performance of embedding vectors of expertise for other problems in software engineering besides the classification of developers in their job roles.
%\appendix

%\section{Sample Appendix Section}
%Lorem ipsum dolor sit amet, consectetur adipiscing elit, sed do eiusmod tempor section \ref{sec:sample1} incididunt ut labore et dolore magna aliqua. Ut enim ad minim veniam, quis nostrud exercitation ullamco laboris nisi ut aliquip ex ea commodo consequat. Duis aute irure dolor in reprehenderit in voluptate velit esse cillum dolore eu fugiat nulla pariatur. Excepteur sint occaecat cupidatat non proident, sunt in culpa qui officia deserunt mollit anim id est laborum.

 \bibliographystyle{elsarticle-num} 
 \bibliography{cas-refs}





% \end{thebibliography}
\end{document}